\documentclass{moriond}

\def\be{\begin{equation}}
\def\ee{\end{equation}}
\def\bea{\begin{eqnarray}}
\def\eea{\end{eqnarray}}

\def\ttbar{\mathrm{t}\overline{\mathrm{t}}}
\def\pt{p_{\mathrm{T}}}

\newcommand{\Photo}{\includegraphics[height=29mm]{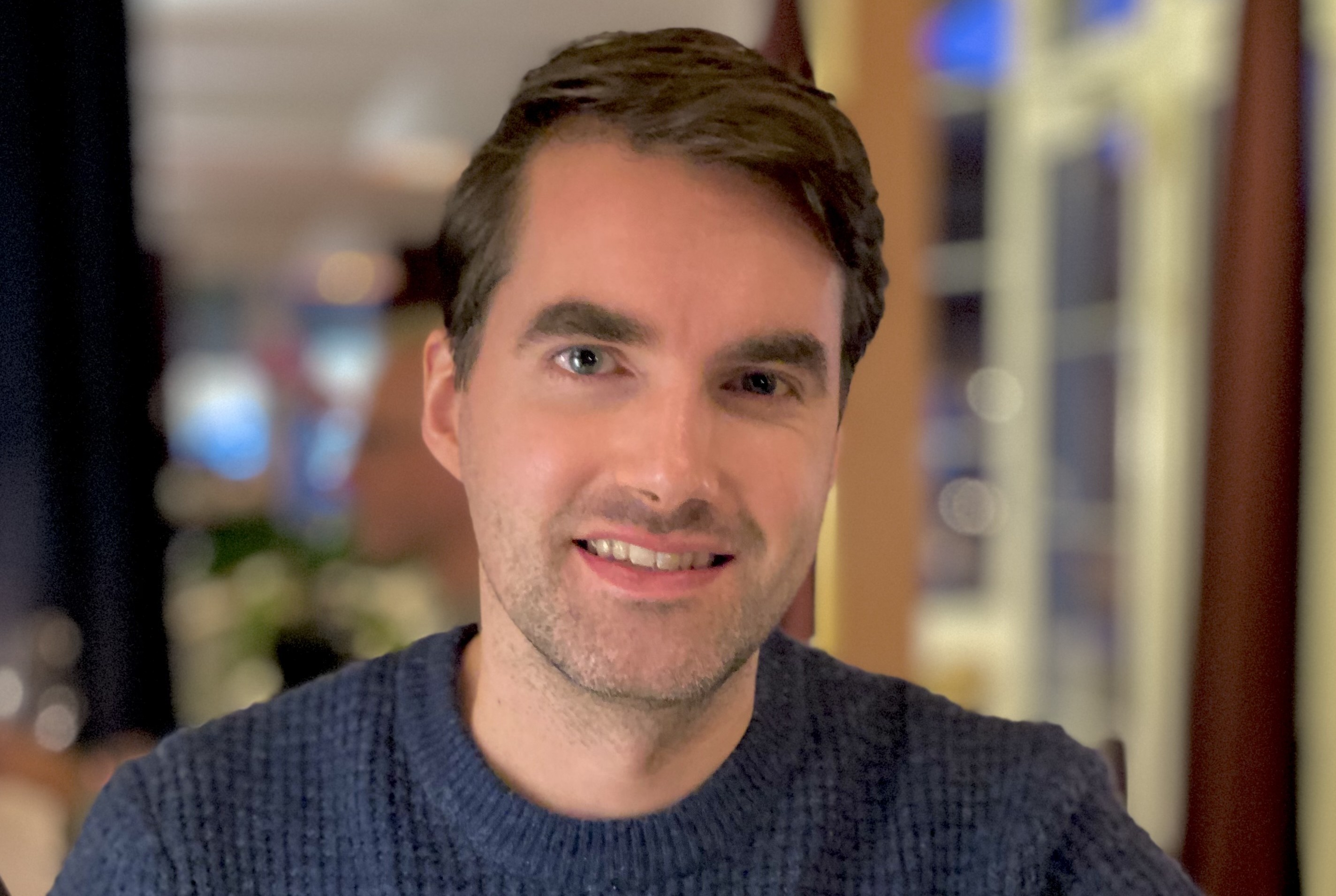}}

\begin{document}
\vspace*{4cm}
\title{FLAVOR CHANGING NEUTRAL CURRENTS AND PROPERTIES \\IN TOP QUARK EVENTS AT THE CMS AND ATLAS EXPERIMENTS}

\author{ GERRIT VAN ONSEM \\ On behalf of the ATLAS and CMS Collaborations}

\address{University of Rochester, Department of Physics and Astronomy,\\
500 Wilson Blvd, Rochester, NY 14627, United States of America
}
\footnotetext{\copyright~Copyright 2023 CERN for the benefit of the ATLAS and CMS Collaborations. Reproduction of this article or parts of it is allowed as specified in the CC-BY-4.0 license.}

\maketitle\abstracts{
The latest searches for flavor changing neutral currents (FCNCs) and measurements of properties in top quark events at the CMS and ATLAS experiments are presented. The vast amounts of proton-proton collision data recorded during the CERN LHC Run 2 at $\sqrt{s}=13$~TeV provide excellent opportunities to test the standard model (SM) with increasingly high precision and search for subtle deviations from the SM expectation. Tight constraints on top quark FCNC decay branching fractions are derived, and asymmetries of observables sensitive to new physics are measured.
}

\section{FCNC and top quark properties as gateways to new physics}
In the standard model (SM) of particle physics, flavor changing neutral current (FCNC) interactions are heavily suppressed, with branching fractions ($\mathcal{B}$) for the decay of top quarks to an up or charm quark and a Z boson, Higgs boson, photon $\gamma$, or gluon not exceeding $10^{-12}$. However, these branching fractions may be enhanced by physics behond the SM by many orders of magnitude.
The FCNC interactions can be probed at the LHC in top quark \textit{production} as well as \textit{decay}. In single top quark production there is a higher sensitivity to the up quark coupling than the charm quark coupling because of the proton parton distribution functions, while in top quark pair ($\ttbar$) production, the sensitivity to these two couplings is more similar.

Another gateway to new physics is provided by precise measurements of top quark properties. At leading order in the SM, the central-forward charge asymmetry $A^{\ttbar}_\mathrm{C}$ in $\ttbar$ events, defined from the difference $\Delta |y|$ between the absolute values of the top quark and antiquark rapidities $y$, is zero. High-order amplitudes cause an asymmetry, resulting in slightly more forward top quarks and more central top antiquarks. 
The presence of new physics may alter this asymmetry and the effect may depend on the kinematic region. 
By measuring asymmetries $A_{\mathcal{O}_i}$ in CP-odd observables $\mathcal{O}_i$ one can also search for CP violation. One possible cause of CP violation would be a chromoelectric dipole moment (CEDM) that modifies the $\ttbar$ production vertex. 
Furthermore, the W boson polarization states can be measured as a precision test of the SM.

\section{Recent results from the CMS and ATLAS experiments}

\subsection{Searches for FCNC interactions in top quark events}

\begin{figure}
\begin{minipage}{0.49\linewidth}
\centerline{\includegraphics[width=0.9\linewidth]{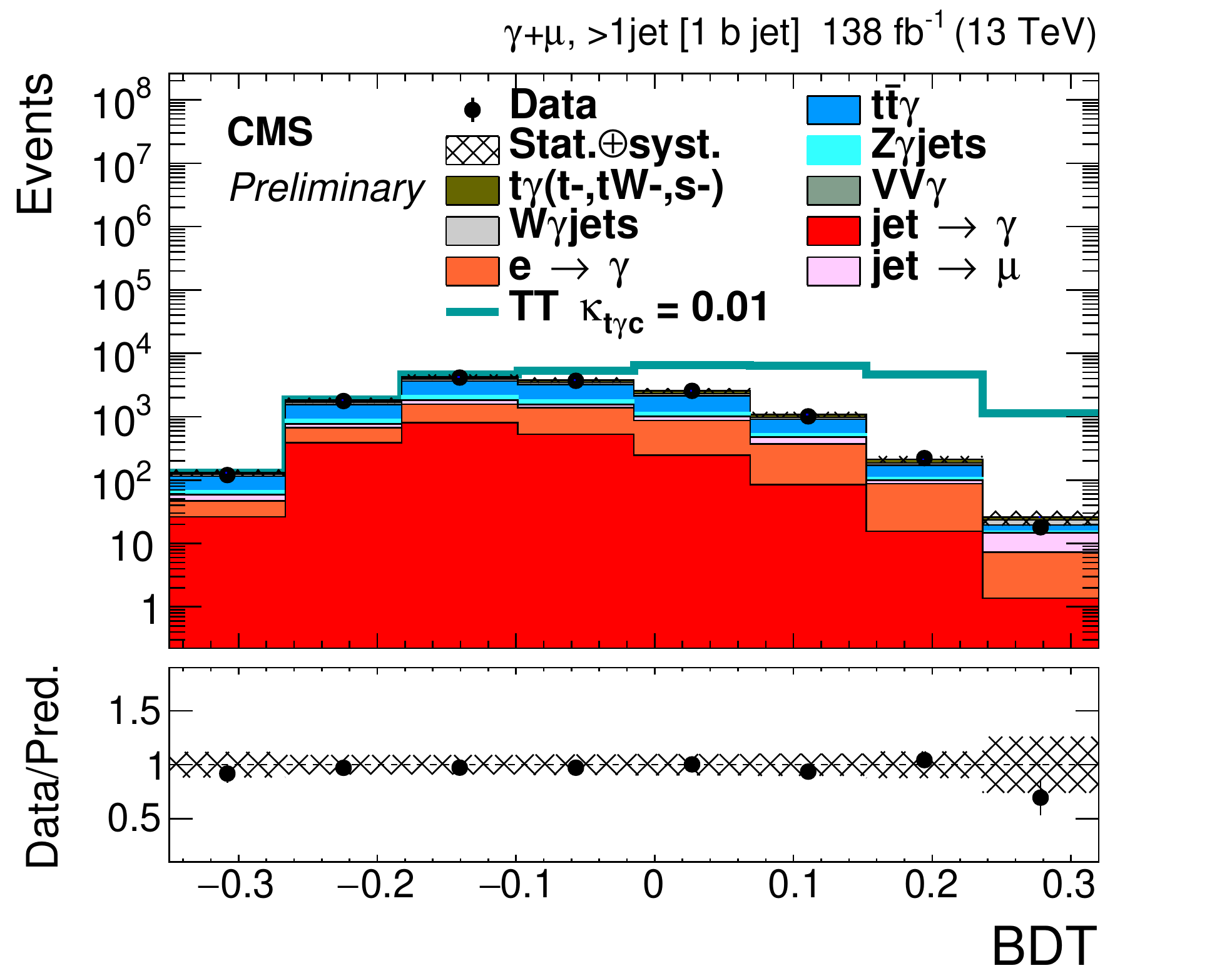}}
\end{minipage}
\hfill
\begin{minipage}{0.49\linewidth}
\centerline{\includegraphics[width=0.75\linewidth]{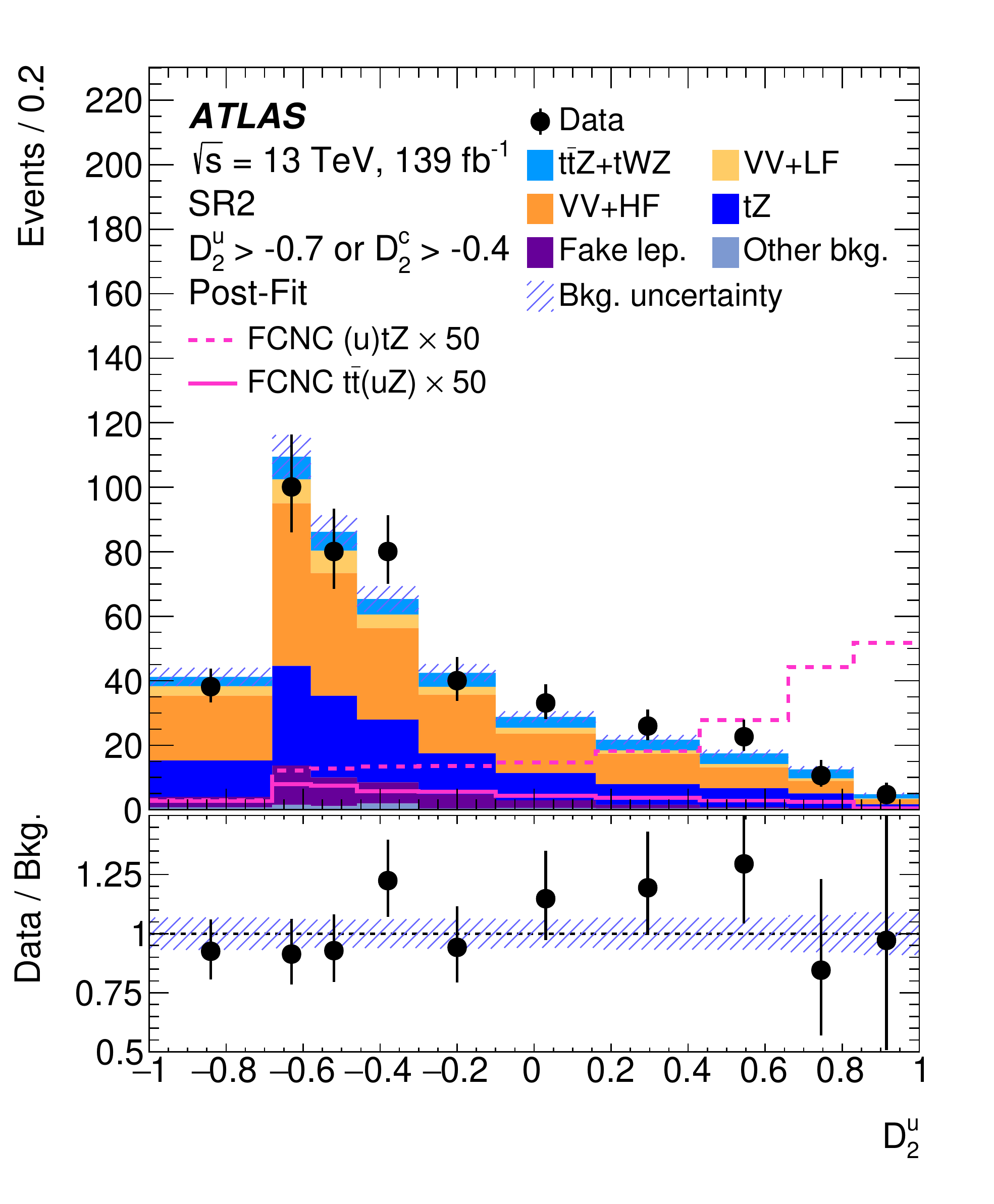}}
\end{minipage}
\caption[]{Example BDT output distributions in searches for FCNC t$\gamma$ interactions by the CMS Collaboration~\cite{TOP-21-013} (left) and for FCNC tZ interactions by the ATLAS Collaboration~\cite{2301.11605} (right). Higher values of the BDT scores typically correspond to more FCNC signal-like events. The observed data are compatible with the SM background prediction.}
\label{fig:FCNC_discriminants}
\end{figure}

In a new search for t$\gamma$ FCNC interactions~\cite{TOP-21-013} at the CMS experiment~\cite{CMS}, one lepton ($\ell = \mathrm{\mu,e}$) and one photon are selected. Two signal regions (SRs) are defined: exactly one b-tagged jet is required for the SR targeting single t production, and at least 2 jets, one of which is b tagged, for the SR targeting $\ttbar$ production.  
To discriminate FCNC signal from SM background, a boosted decision tree (BDT) is trained, with 13 input features, such as the $\pt$ and $\eta$ of the photon and lepton, the lepton charge, and angular separation $\Delta R$ between objects. An example distribution of the BDT output is shown in Fig.~\ref{fig:FCNC_discriminants} (left).
Uncertainties in this measurement include uncertainties in the fake leptons and photon background estimation, and the photon energy scale. 
Via a simultaneous fit of the BDT distributions in the two SRs, upper limits at $95\%$ CL on the FCNC branching fractions of $\mathcal{B}(\mathrm{t} \rightarrow \mathrm{u\gamma)} < 0.95 \times 10^{-5}$ and $\mathcal{B}(\mathrm{t} \rightarrow \mathrm{c\gamma}) < 1.51 \times 10^{-5}$ are derived. These results provide a signficant improvement with respect to the previous CMS result, partly thanks to the addition of the e channel and new SRs.

A search was performed at the ATLAS experiment~\cite{ATLAS} for FCNC tH interactions with the Higgs boson decaying to two $\tau$ leptons~\cite{2208.11415}. 
Events are categorized into leptonic and hadronic channels targeting specific decay modes of the top quark and the Higgs boson in single t or $\ttbar$ production. A multitude of SRs and control regions (CRs) are defined based on the number of jets, b jets, light leptons, and hadronically decaying $\tau$ leptons. 
In each SR, BDTs are designed, combining 12--17 input features such as the $\pt$ of the leading $\tau$ lepton, and the invariant mass of the $\tau$ lepton pair. 
Dominant uncertainties include the statistical uncertainty in the data, the limited size of the simulated samples, and fake $\tau$ modeling.
A simultaneous fit of the BDT output distributions in all SRs reveals a slight excess of data of about two standard deviations, and upper limits are set on the FCNC branching fractions: $\mathcal{B}(\mathrm{t} \rightarrow \mathrm{uH}) < 6.9 \times 10^{-4}$ and $\mathcal{B}(\mathrm{t} \rightarrow \mathrm{cH}) < 9.4 \times 10^{-4}$.

A search for FCNC tZ interactions was performed at the ATLAS experiment, using events with three charged leptons~\cite{2301.11605}.
Two SRs are defined: in the first, targeting FCNC in $\ttbar$ decays, at least 2 jets are required, and in the other region, targeting FCNC in single t production, 1 or 2 jets are required. 
The top quark reconstruction is performed via $\chi^2$ methods in specific FCNC hypotheses.
In each SR, a BDT is trained with 6 input features, such as the mass of the SM and FCNC decaying top quark candidates, and $\Delta$R between reconstructed top quark candidates. An example BDT distribution in the SR targeting single top quark production is shown in Fig.~\ref{fig:FCNC_discriminants} (right).
Leading uncertainties are statistical, and systematic uncertainties in the tZ normalization and diboson modeling. 
A simultaneous fit of BDT distributions and event yields in SRs and CRs is performed to extract the bounds on the branching fractions for left-handed ($\mathcal{B}(\mathrm{t} \rightarrow \mathrm{uZ}) < 6.2 \times 10^{-5}$ and $\mathcal{B}(\mathrm{t} \rightarrow \mathrm{cZ}) < 13 \times 10^{-5}$) and right-handed ($\mathcal{B}(\mathrm{t} \rightarrow \mathrm{uZ}) < 6.6 \times 10^{-5}$ and $\mathcal{B}(\mathrm{t} \rightarrow \mathrm{cZ}) < 12 \times 10^{-5}$) couplings separately.
These results improve on the previous ATLAS result because of the addition of the single t channel, the BDTs, and the larger data set.

\subsection{Measurements of $t\overline{t}$ charge asymmetry}
The CMS Collaboration measured the charge asymmetry in single lepton $\ttbar$ events with boosted top quarks~\cite{2208.02751}.
Three classes of topologies are defined based on the number of top quark and W boson tags from large-angle jet tagging: boosted, semiresolved, and resolved events. 
The top quark pair reconstruction is performed by assigning top- or W-tagged jets to hadronic or leptonic legs of the $\ttbar$ decay. 
Analysis channels are defined by the range of the $\ttbar$ invariant mass ($m_{\ttbar}$), year, and $\ell$ flavor.
The unfolded charge asymmetry is obtained from a simultaneous fit across these channels of event yields with positive and negative $y$ differences. 
Statistical uncertainties and systematic uncertainties in the $\ttbar$ modeling such as matrix element and parton shower scales are dominant. 
The charge asymmetry is measured to be $A^{\ttbar}_\mathrm{C} = (0.69^{+0.65}_{-0.69})\%$, in agreement with NNLO SM calculations but also still compatible with 0, and is also derived as a function of the $m_{\ttbar}$ range, indicating the compatibility between the data and the prediction.

\begin{figure}
\begin{minipage}{0.49\linewidth}
\centerline{\includegraphics[width=0.94\linewidth]{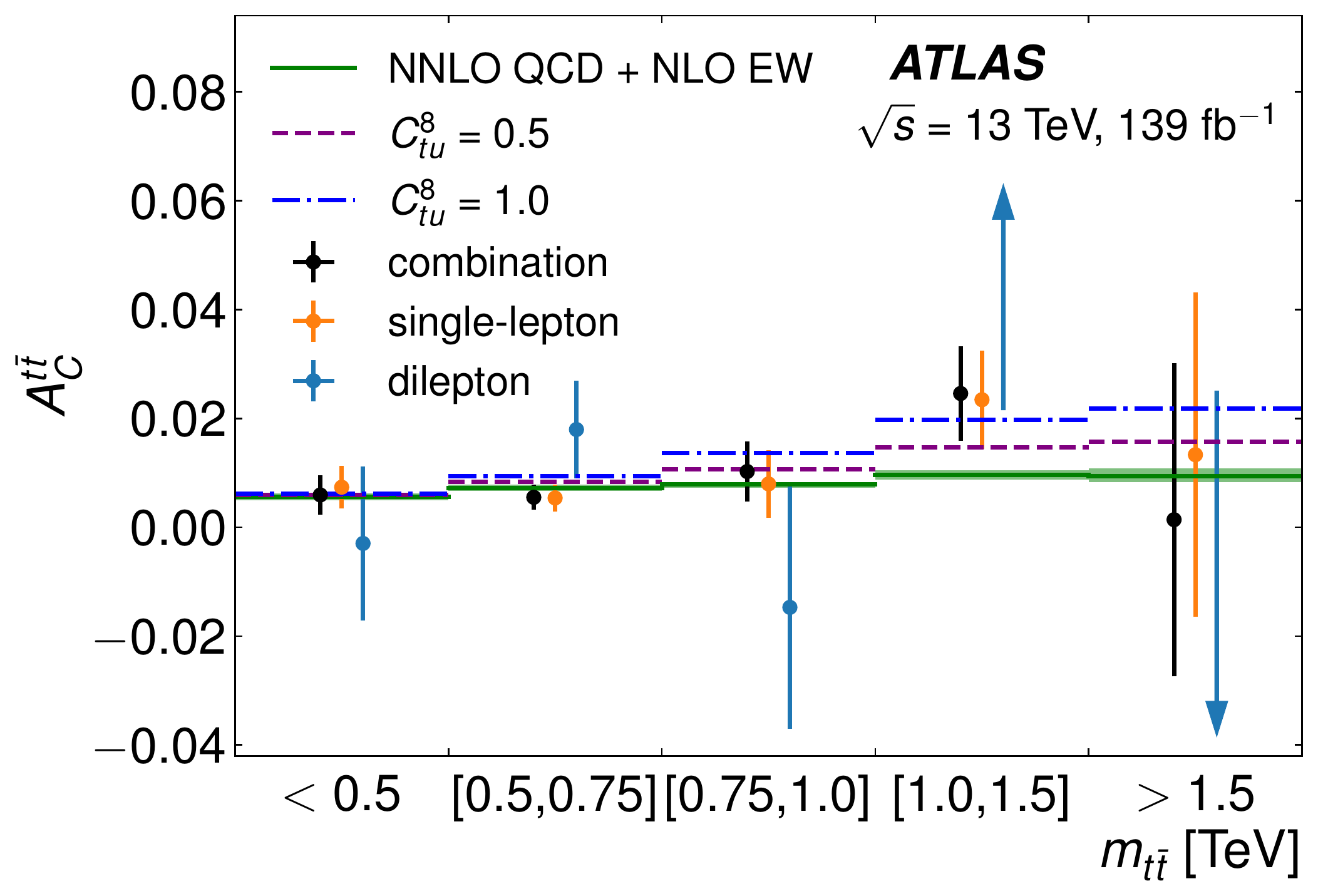}}
\end{minipage}
\hfill
\begin{minipage}{0.49\linewidth}
\centerline{\includegraphics[width=0.89\linewidth]{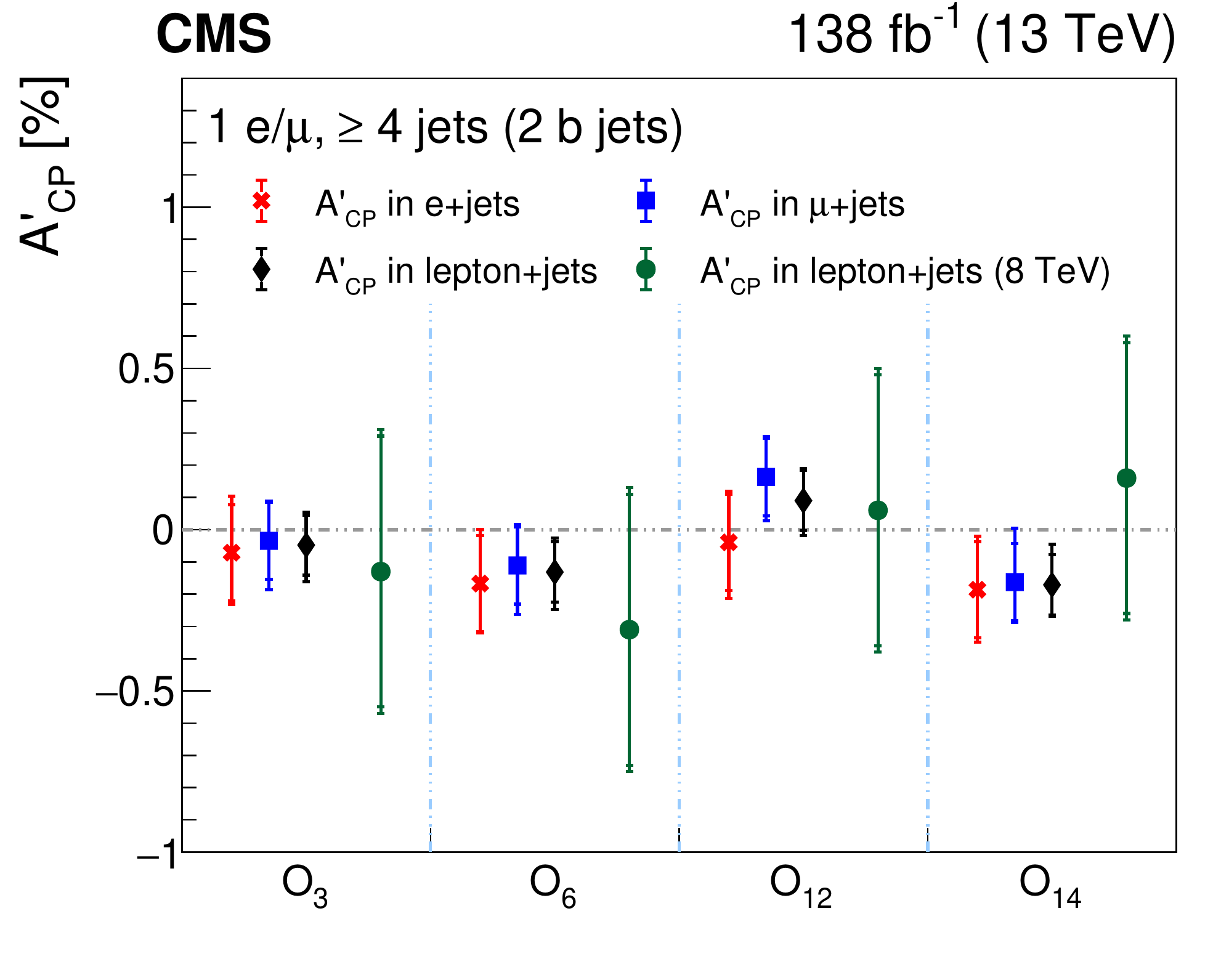}}
\end{minipage}
\caption[]{The charge asymmetry as function of $m_{\ttbar}$ measured at the ATLAS experiment~\cite{2208.12095} (left) and the measured asymmetries in CP-odd observables in $\ttbar$ events at the CMS experiment~\cite{2205.02314} (right) are compatible with the SM.}
\label{fig:chargeasymmetry_CPviolation}
\end{figure}

The charge asymmetry in single lepton and dilepton $\ttbar$ events was also measured at the ATLAS experiment~\cite{2208.12095}.
The $\ttbar$ reconstruction is done differently in each of three classes of events. In the single lepton resolved topology, jets are assigned to the $\ttbar$ hypothesis via a BDT separating correct and wrong jet combinations. In the single lepton boosted topology, the leading large-angle jet is assigned as the hadronically decaying top quark. Finally, in the dilepton channel, only looking at resolved topologies, the reconstruction is performed via a neutrino weighting method.
The leading uncertainties are statistical and systematic uncertainties in the $\ttbar$ modeling. 
The $\Delta |y|$ distributions are unfolded inclusively, leading to a charge asymmetry of $A^{\ttbar}_\mathrm{C} = (0.68 \pm 0.15)\%$, 4.7 standard deviations away from 0, therefore providing strong evidence of charge asymmetry in $\ttbar$ events.
It is also unfolded as a function of several kinematic observables, such as $m_{\ttbar}$, shown in Fig.~\ref{fig:chargeasymmetry_CPviolation} (left), showing compatibility with the SM prediction. 

\subsection{Searches for CP violation in $t\overline{t}$ events}
The CMS Collaboration searched for CP violation in dileptonic $\ttbar$ events~\cite{2205.07434}. 
Two CP-odd Lorentz scalars, $\mathcal{O}_1$ and $\mathcal{O}_3$, are constructed from four-momenta of the leptons and the reconstructed top quarks, and the leptons and the b quarks originating from the top quark, respectively. 
The top quark pair is reconstructed by solving a system of equations for the neutrino momentum, using W boson mass and top quark mass constraints. 
The leading uncertainties are statistical, and systematic uncertainties in $\ttbar$ modeling, such as color reconnection. 
A fit is performed to extract simultaneously the asymmetry and the $\ttbar$ cross section from positive and negative regions of the CP-odd observable under consideration, and the results agree with the SM prediction of no CP violation: $A_{\mathcal{O}_1} = (2.4 \pm 2.8) \times 10^{-3}$ and $A_{\mathcal{O}_3} = (0.4 \pm 2.8) \times 10^{-3}$. 
The asymmetry is proportional to the CEDM parameter $d_\mathrm{tG}$, and the measurement is translated into bounds on the CEDM.  

The search for CP violation in $\ttbar$ production has also been performed at the CMS experiment in events with a single lepton~\cite{2205.02314}.
Four CP-odd observables,  $\mathcal{O}_i$, are calculated from four-momenta of reconstructed objects. 
Statistical uncertainties and systematic uncertainties in the $\ttbar$ modeling such as the matrix-element parton-shower matching are dominant. 
The dilepton contribution in the selected events may cause a spurious asymmetry measurement. Therefore the single $\ell$ signal and dilepton background yields in positive and negative regions of a particular CP-odd observable is extracted via a fit to the $m(\ell,b)$ distribution. The measured asymmetries for each CP-odd observable is shown in Fig.~\ref{fig:chargeasymmetry_CPviolation} (right), split by lepton channel and combined.
The dimensionless CEDM is measured to be $d_\mathrm{tG} = 0.04 \pm 0.10 ~\mathrm{(stat)} \pm 0.07~\mathrm{(syst)}$, agreeing with the SM prediction of no CP violation.

\subsection{Measurement of the W boson polarization}
Finally, a measurement of the W boson polarization in top quark decays in dilepton events by the ATLAS Collaboration is presented~\cite{2209.14903}.
The $\ttbar$ reconstruction is performed via a neutrino weighting algorithm. 
The dominant uncertainties are in the $\ttbar$ modeling, the jet energy scale and resolution, and lepton reconstruction.
The polarization is extracted from the unfolded distribution of the cosine of the $\theta^*$ angle, which is the angle between the $\ell$ momentum from the W boson decay and the reversed direction of the b quark from the top quark decay, both measured in the W boson rest frame. The differential cross section is fitted with an analyticial function to obtain the longitudinal, left-handed, and right-handed W boson helicity fractions. The measurements agree with the SM NNLO prediction, and the experimental uncertainty is now nearly at the level of the theory uncertainty for the left-handed helicity fraction.

\section{Summary and outlook}
The ATLAS and CMS Collaborations continue to analyze the vast amounts of top quark event data of the full LHC Run 2. 
There is no sign of new physics, yet, and SM predictions are confirmed with unprecedented precision. 
Results from many more Run 2 measurements are in the pipeline. Moreover, as Run 3 is well underway and will double the Run 2 data set, the next years provide exciting prospects in the search for new physics in the top quark sector.

\end{document}